\documentstyle[12pt,epsf]{article}
\def\reference{\parskip 0pt\par\noindent\hangindent 0.5 truecm}
\def\kms{km ${\rm s}^{-1}$}

\def\la{\mathrel{\hbox{\rlap{\hbox{\lower4pt\hbox{$\sim$}}}\hbox{$<$}}}}
\def\ga{\mathrel{\hbox{\rlap{\hbox{\lower4pt\hbox{$\sim$}}}\hbox{$>$}}}}
\def\arcmin{\hbox{$^\prime$}}

\def\fm{\hbox{$.\!\!^{\rm m}$}}
\def\mag{\hbox{$^{\rm m}$}}

\def\farcm{\hbox{$.\mkern-4mu^\prime$}}

\def\deg{{^\circ}}
\def\hi{\mbox{\normalsize H\thinspace\footnotesize I}}
\newcommand{\etal}{{\it et al.}\,}      
\newcommand{\eg}{{\it e.g.},\ }         
\newcommand{\ie}{{\it i.e.},\ }         
\newcommand{\cf}{{\it cf.},\ }          
\begin{document}
%
%
\title{Large-Scale Structures Behind the Southern Milky Way 
from Observations of Partially Obscured Galaxies}
\author
 {R.C. Kraan-Korteweg\\Observatoire de Paris-Meudon, D.A.E.C.,
  92195 Meudon Cedex, France\\kraan@gin.obspm.fr\\
\and 
 P.A. Woudt\\Dept. of Astronomy, University of Cape Town,
 Rondebosch 7700, South Africa\\
\and 
P.A. Henning\\Dept. of Physics and Astronomy, University of
     New Mexico, Albuquerque, USA
}
\date{} 
\maketitle

\begin{abstract}
We report here on extragalactic large-scale structures uncovered by
a deep optical survey for galaxies behind the southern Milky Way. 
Systematic visual inspection of the ESO/SRC-survey revealed over 
10000 previously unknown galaxies in the region $265\deg \la \ell 
\la 340\deg, |b| \la 10\deg$. With subsequently obtained redshifts
of more than 10\% of these galaxies, new structures across the Milky
Way are unveiled, such as a filament at $v\sim 2500$ \kms\
connecting to the Hydra and Antlia clusters, a shallow extended 
supercluster in Vela ($\sim 6000$ \kms), and a nearby (4882 \kms), 
very massive (${\cal M}\sim 2-5 \cdot10^{15}{\cal M_\odot}$),
rich Coma-like cluster which seems to constitute the previously
unidentified center of the Great Attractor.  

The innermost part of the Milky Way where the foreground 
obscuration in the blue is ${\rm A}_B \ga 5\mag$, respectively \hi-column 
densities ${\rm N}_{\hi} \ga 6\cdot 10^{21} {\rm cm}^{-2}$ remains
fully opaque. In this approximately $8\deg$ wide strip, the 
forthcoming blind \hi-survey with the multi-beam system at Parkes will
provide the only tool to unveil this part of the extragalactic sky.
\end{abstract}


\bigskip
\section{Introduction}
The Milky Way obscures about 25\% of the extragalactic sky. This severely 
constrains studies of:
\begin{itemize}
\item 
Large-scale structures, particularly the connectivity of the 
Supergalactic Plane, other superclusters, walls and voids across
the Milky Way.
\item
The origin of the peculiar motion of the Local Group (LG) with 
respect to the Cosmic Microwave Background Radiation (CMB). Can the dipole
in the CMB be explained by the gravity  of the irregular
mass/galaxy distribution in the {\sl whole} sky?
\item 
Other streaming motions. Is the predicted mass overdensity, the Great
Attractor (GA) -- as evidenced in a large-scale systematic flow of galaxies
towards  ($\ell,b,v)\sim(320\deg,0\deg,4500$ \kms) (Kolatt et
al. 1995) -- in the form of galaxies, hence does light trace mass?
\item
Individual nearby galaxies. Could a nearby Andromeda-like
galaxy lie hidden in the Zone of Avoidance (ZOA) -- important  
for the internal dynamics of the LG, as well as   
mass derivations of the LG and the present density of the Universe 
from timing arguments (Peebles 1994). Moreover, the gravitational 
attraction of the nearest galaxies ($v<300$ \kms) 
generate 20\% of the total dipole moment (Kraan-Korteweg 1989). 
\end{itemize}
In recent years various groups have initiated projects to unveil
the galaxy distribution behind our Milky Way. The methods are
manifold (\cf {\it Unveiling of Large-Scale Structures behind the Milky Way}, 
1994, for a review). 
Here, we describe the results from a deep optical survey for galaxies
in the southern Milky Way, a particularly interesting region
because of the dipole in the CMB and the infall into the GA
both of which point close to the southern Galactic Plane (\cf Fig.~1).

\section{A Deep Optical Galaxy Search in the Southern ZOA} 
To reduce the $\sim 20 \deg$-wide strip of the ZOA, we have embarked 
on a deep search for partially obscured galaxies, \ie down to fainter
magnitudes and smaller dimensions compared to existing catalogues. 
To date, 50 fields of the ESO/SRC-survey have been visually
inspected. The surveyed area lies within $265\deg \la \ell \la
340\deg, |b| \la 10\deg$ -- its borders are outlined
in Fig.~1.  
Within the surveyed area of $\sim 1200\Box\deg$, 10276 new 
galaxies with ${\rm D}\ge 0\farcm2$ have been identified in addition to the 
269 Lauberts galaxies with ${\rm D} \ge 1\arcmin$ within 
this area (Lauberts 1982). Their distributions are displayed in Fig.~1.
\begin{figure}[h]
\hfil \epsfxsize 17.5cm \epsfbox{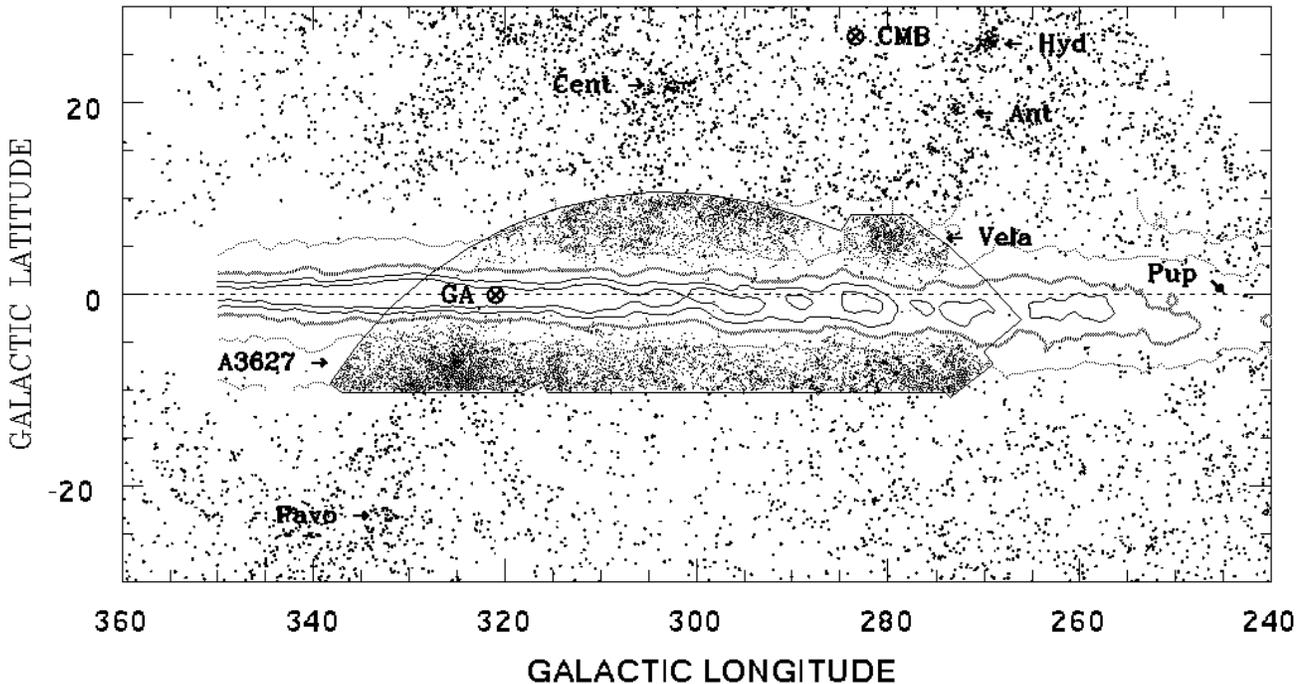}\hfil
\caption
{The galaxy distribution from our deep galaxy search (${\rm D} \ge
0\farcm2$) and the Lauberts galaxies (${\rm D} \ge 1\arcmin$) in 
Galactic coordinates. The to date surveyed region is outlined. 
The dipole direction in the CMB, 
the center of the GA and the most prominent clusters are marked.
The contours mark absorption in the blue of ${\rm A}_B$ =  $1\mag,
2\fm5, 5\mag$ (thick contour), $7\fm5, 10\mag$ as derived from the Galactic
HI (Kerr \etal 1986), adopting a constant gas-to-dust ratio
and the formalism given by Burstein \& Heiles 1982.}
\end{figure}
Details on the search can be found in Kraan-Korteweg \& Woudt 1994.

In the mean, the galaxy density is well
correlated with the foreground extinction A$_B$ as traced
by the HI-column-density: for ${\rm A}_{B} \ga 5\mag$ (thick contour), respectively
${\rm N}_{\hi}\ga 6\cdot 10^{21} {\rm cm}^{-2}$, the ZOA remains opaque. 
Above this band, distinct filaments and round concentrations
uncorrelated with the foreground dust can be recognized, thus
they must have their origin in extragalactic
large-scale structures. Some of these features seem 
to align with the known galaxy distribution, as \eg a filament above
the Galactic Plane which points toward the Centaurus cluster, 
and the filament from the Hydra and Antlia clusters towards the 
prominent overdensity in Vela. 
A significant overdensity, centered on the previously identified
cluster A3627 (Abell \etal 1989), is evident within only a few degrees
of the predicted center of the GA. 
It is the only Abell cluster behind the Milky Way, classified as rich 
and nearby and lies within the GA region. Even so, this 
cluster has received little attention because of the 
diminishing effects of the foreground obscuration. 
And this cluster at $(\ell,b) = (325\deg,-7\deg$) is hardly obvious in
the distribution of Lauberts galaxies and not at all in IRAS samples.
However, the galaxies in this overdensity are on average relatively large
(just {\sl below} the diameter limit of Lauberts). 
Lifting the obscuring veil of the Milky Way 
by correcting the observed properties of the galaxies for 
the foreground extinction would, in fact, reveal this cluster as the
most prominent overdensity in the southern sky.

Redshift measurements are required to map the galaxies in 3 dimensions.
The 2-dimensional galaxy distribution alone can be misleading.
For instance, the prominent overdensity in Vela was found to be due to
a superposition of a nearby ($\sim 2500$ \kms) filament connecting to
Hydra, a more distant (6000 \kms) shallow extended supercluster, and
a very distant ($\sim 16000$ \kms) wall-like structure crossing the ZOA
(Kraan-Korteweg \etal 1994). 

\section{\hi-Observations of Obscured Galaxies at Parkes}
The redshifts of the uncovered galaxies are obtained by three
observational approaches:
(a) multifiber spectroscopy with OPTOPUS and MEFOS at the 3.6m telescope 
of ESO in the densest regions, (b) individual spectroscopy of the
brightest galaxies with the 1.9m telescope of the SAAO, and (c) 21cm 
observations of extended low surface-brightness (LSB) spirals with
the 64m Parkes radio telescope. So far we have obtained 1083 new redshifts
in our search area. 

An extensive discussion of methods and typical results are given in
Kraan-Korteweg \etal 1994. The three
observing methods are complementary in galaxy populations,
characteristic magnitude and diameter range and the depth of
volume they probe. The multifiber spectroscopy gives a good
description of clusters and dense groups in the ZOA and traces 
large-scale structures out to recession velocities of 25000
\kms. The SAAO and \hi-observations cover the bright end of 
the galaxy distribution and provide a more homogenous sampling of
galaxies out to 10000 \kms. 
 
Our \hi-observations at Parkes have similar sensitivities as the
forthcoming blind HI-survey with the multi-beam (MB) receiver in the ZOA
and therefore merit a more detailed description.
These observations are vital in recovering an important fraction of the 
nearby spiral galaxy population which would otherwise be impossible to map.
Although we aimed at complete coverage for the brightest galaxies with
the optical spectroscopy, a significant fraction of apparent bright galaxies
cannot be traced in this manner. In general, this concerns nearby 
spirals (and also dwarfs) which -- seen through a layer of extinction 
-- are extended, very LSB objects.

We have so far observed 345 spiral galaxies without previous redshift
estimates at Parkes. Each galaxy was observed in total power mode, a 
bandwidth of 32 MHz and the 1024 channel autocorrelator. We generally 
integrated 30 minutes ON / 30 minutes OFF source. This yielded an
r.m.s. after Hanning smoothing of typically 4 mJy -- hence comparable in
sensitivity to the proposed MB ZOA-survey. However, we often 
detected strong galaxies in as little as ten minutes.  
On the other hand, we sometimes observed a source for an hour
to get sufficient signal to noise. We found the sensitivity during 
the day to be significantly degraded compared to the level achieved at
night due to high-amplitude standing waves in the spectrum which
precluded detections even with a large number of very short
integrations.

In 1993 we generally centered the recession velocity at 
3000 \kms\ and reobserved some of the non-detections at a central velocity of 
7500 \kms\ in a second step. This resulted in a detection
rate of over 50\%. In 1994 and 1995, we used two IF's and offset 
512 channels of each polarization by 22 Mhz. This resulted in a
velocity coverage of 0-10000 \kms\ with the 32 MHz in one
integration. Although the lower frequencies were often badly disturbed
by interference around 8300 \kms\ (\cf Fig.~2), this increased our
detection rate to nearly 80\% in 94! Our observations were
considerably less effective in the 95-run due to a strong increase in 
recurring interference, part of which were generated locally by the 
then ongoing tests of the pulsar equipment -- a problem that demands 
careful investigation of the Galactic pulsar-survey which will be piggy-backed 
on the MB ZOA-survey!

A few examples of typical HI-spectra, 
recurring interference as well as detections at low Galactic
latitudes are illustrated in Fig.~2.
\begin{figure}[h]
\hfil \epsfxsize 17.5cm \epsfbox{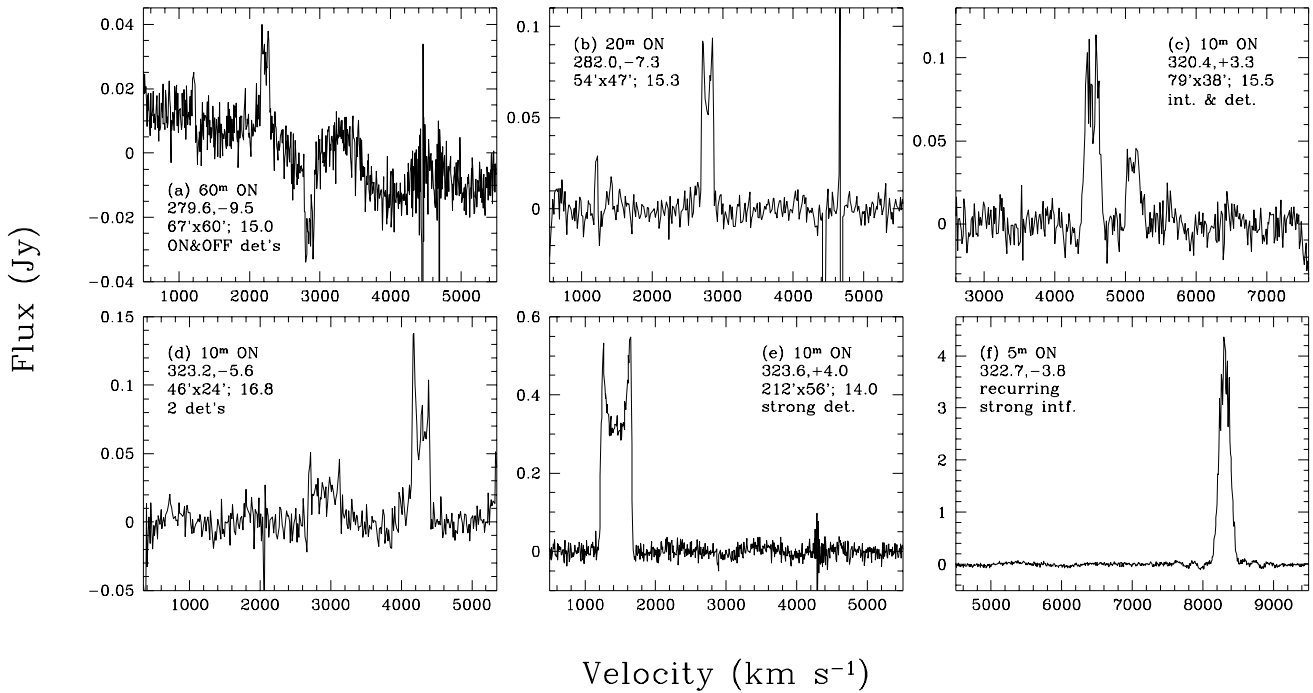}\hfil
\caption
{Typical spectra of galaxies in the ZOA obtained with the 64m Parkes 
radiotelescope. Integration times (between 5$^{\rm min}$ and 60$^{\rm min}$), 
Galactic coordinates, observed
(absorbed) diameters (in arcsec) and magnitudes in the blue B$_J$ are 
indicated. Except for panel {\it a}, all spectra are
baseline-subtracted. Regularly appearing interference of different 
strengths are found at 1200, 4450, 4700 and 8300 \kms.}
\end{figure}
Panel {\it a} and {\it b} are pointed observations in the Hydra/Antlia
extension, {\it c - f} are observations in the GA region. Panel {\it
a} shows a detection in the ON as well as a detection in
the adjacent $10\fm5$ earlier OFF-position in the crowded 
nearby Hydra/Antlia-filament. Panel {\it b} shows another 
example of a previously unidentified LSB member of this filament.
Note the recurring interference at 1200, 4450 and 4700 \kms. 
The interference at 1200 \kms\ is found in practically all
scans (positive or negative). The final shape often 
resembles real \hi-profiles which may be interpreted as erroneous
detections. Moreover, detections of weaker real galaxies within that 
velocity range are impossible. 

Panel {\it d} shows 2 detections within one beam in the densely
populated GA-region, whereas in panel {\it c} only the feature at 
5100 \kms\ results from the detection of a massive GA-galaxy; the 
stronger signal at 4450 \kms\ is due to interference.

The interference at 8300 \kms\ (panel {\it f}) constitutes a true
problem. Unstable in time and strength, it has perturbed 
about half of the observations in the higher velocity range. 
Its strength -- sometimes over 10 Jy -- causes ringing and baseline 
wiggles and precludes detections of galaxies in a broad velocity range. This
makes an analysis of the detection rate statistics and especially
the volume completeness function of the MB ZOA-survey extremely hard.

Still, it can be maintained that the \hi-observations recover obscured 
galaxies deeper in the obscuration layer compared to optical 
spectroscopy; a natural extension into the fully obscured region 
will be the MB ZOA-survey.
The relatively high detection rate of obscured low-latitude
galaxies ($5\deg \la |b| \la 10\deg$) forecasts quite a success rate
for the MB ZOA survey ($|b| \le 5\deg$) -- with more than one
detection per beam in high-density areas (\cf Fig.~2{\it a} and {\it d}). 

\section{Uncovered Structures and the Cluster Abell 3627}
The new redshifts along with published redshifts in adjacent regions evidence 
the following structures: 
 
The Hydra and Antlia clusters are not isolated clusters, but
part of a filamentary structure which can be traced from Hydra   
through Antlia (\cf Fig.~1), across the Galactic Plane to
$(\ell,b)=(283\deg,-10\deg$), thus constituting a major structure 
in the nearby Universe ($>35\deg$ at a median
recession velocity of only 2800 \kms).
It seems more a filamentary structure, consisting of spiral-rich
groups and clusters, rather than a supercluster.
 
The prominent galaxy overdensity in Vela is part of a previously
unrecognized shallow, large-scale, overdensity centered on 
$\sim 6000$ \kms. The independent predictions of a supercluster
at this position and distance by Hoffman (1994) and Saunders 
\etal (1991) indicate that it could be quite massive.

The redshifts in the cluster A3627 ($<v>=4882$ \kms\ for 131 reduced to
date) put this cluster near the center of the Great Attractor in 
velocity space. The mass estimate from its velocity dispersion 
($\sigma = 903$ \kms) is that of a rich cluster ($5 \cdot 10^{15}
{\cal M_\odot}$). Its other cluster properties -- predominance of 
early-type galaxies at the center, its core radius R$_c = 10\farcm4$ 
and its central density ${\rm N}_\circ = {\rm 800 gal}/\Box\deg$ all are
consistent with this being a rich massive cluster (Kraan-Korteweg
\etal 1996). It even has, 
like Coma, two dominant central cD-galaxies (\cf Fig.~3).

Rich massive clusters generally are strong X-ray emitters and 
were identified early on with the X-ray satellites (Einstein, HEAO,
Uhuru). Despite dedicated searches and the fact that the hard X-ray 
band is hardly affected by \hi-absorption, 
A3627 had never been identified in X-ray (Jahoda \& Mushotzky
1989, Lahav \etal 1989). 
We therefore studied the ROSAT PSPC data of this cluster (B\"ohringer
\etal 1996) and found that A3627, with an X-ray luminosity of 
L$_X=2.2\cdot 10^{44} {\rm erg s}^{-1}$, is in fact the $6^{th}$-brightest 
X-ray cluster in the ROSAT All Sky Survey. Moreover, the independent 
mass estimate from the X-ray data is consistent with the virial mass.

Fig.~3 displays the X-ray contours (B\"ohringer \etal 1996)
superimposed on an image of the central part of the cluster.
\begin{figure}[h]
\hfil \epsfxsize8.cm \epsfbox{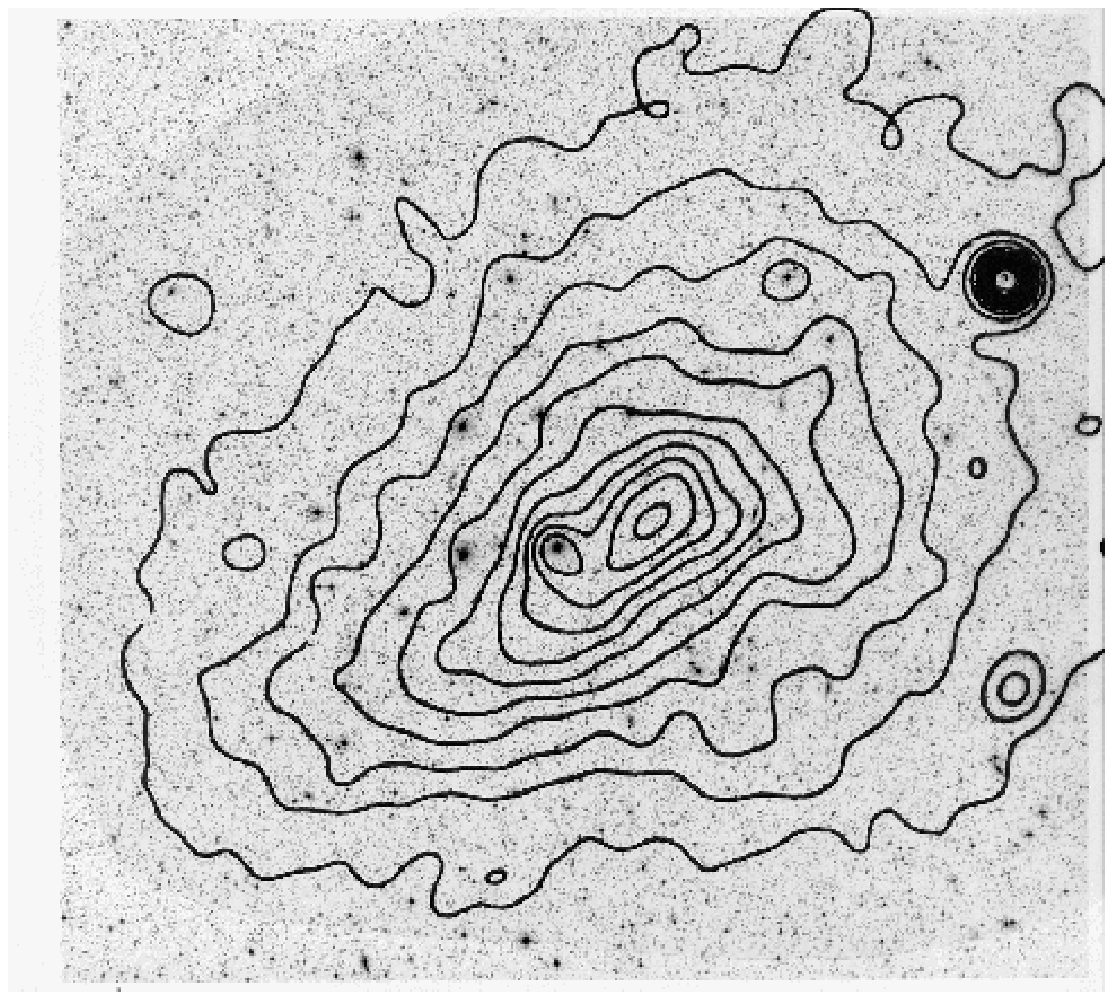} \hspace{0.5cm} 
\epsfxsize8.cm \epsfbox{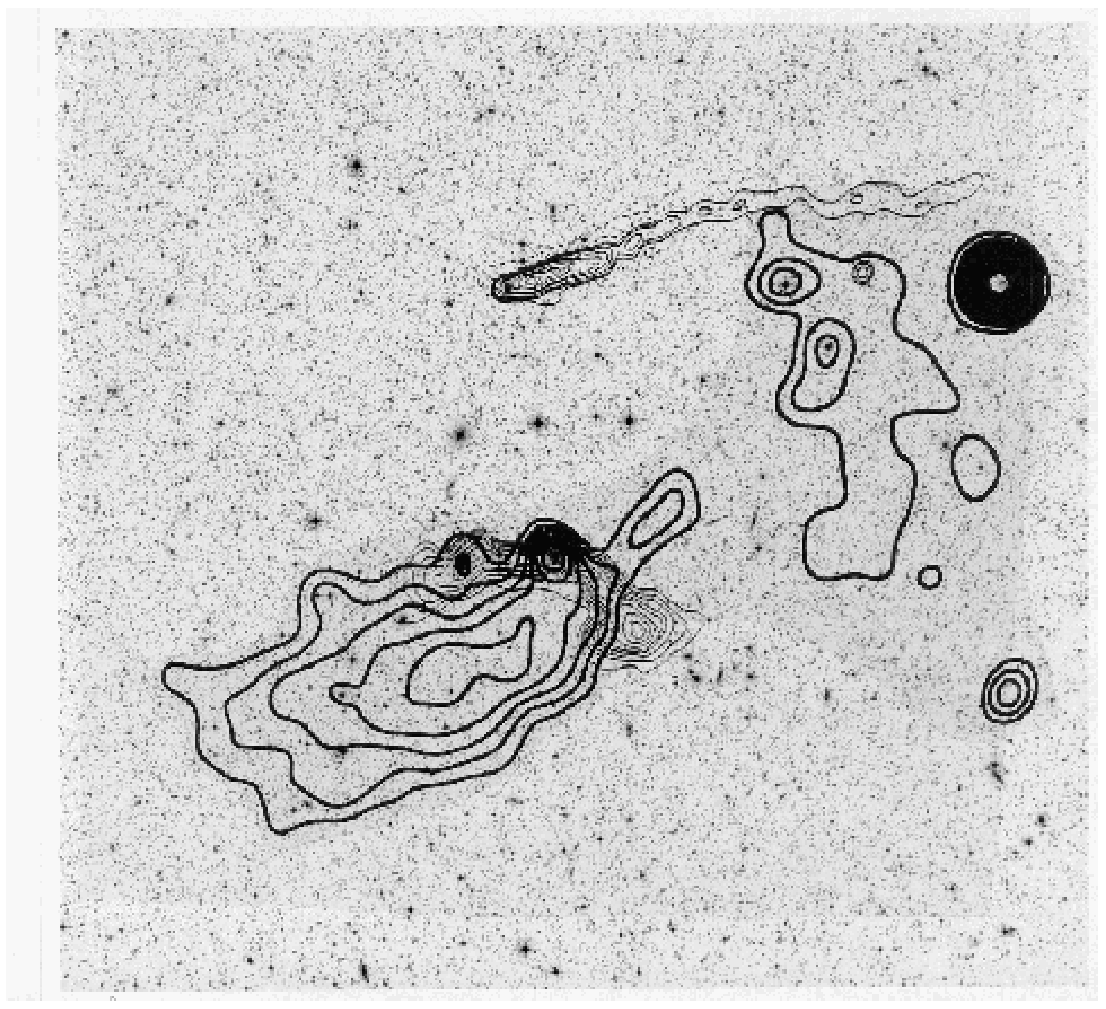} \hfil
\caption
{The central part (56\arcmin x 56\arcmin) of the cluster A3627 as 
reproduced from 
field 136 of the IIIaJ copy of the ESO/SRC survey. Superimposed in the
left panel are the X-ray contours from the ROSAT PSPC
observations; the right panel shows the residual
X-ray contours after substraction of a spherical component,
and the 843 MHz radio continuum emission of the 
wide-angle tail radio galaxy PKS1610-60.8 and the head-tail 
radio-source B1610-60.5. The strong X-ray point-source (top right
corner), a cluster galaxy, is a Seyfert 1 (Woudt \etal 1996).}  
\end{figure}
The X-ray provides interesting insight into the cluster morphology.
The X-ray center is offset from the center of the cluster, the
strong radio source PKS1610-60.8, one of the two cD galaxies. The
latter coincides with the second central X-ray peak. The 
X-ray emission is not symmetric but extended towards the 
SE (left-bottom) corner. Subtracting a spherical symmetric model 
leaves a residual component (\cf right panel).
This subcluster contains the wide-angle-tail radio galaxy 
PKS1610-60.8 whose large radio lobes of 8 arcmin ($\sim 210$ kpc)
have a bending angle of $45\deg$. The contours of the radio emission
(Jones \& Mc Adam 1992) are drawn onto the X-ray subgroup as well. 
Note the alignment of the subclump with the radio emission. The
redshift data are yet too sparse to allow a kinematical analysis of 
this cluster, however, the redshifts in the subclump tend to be 
higher, suggesting that this subgroup is in front of 
the main cluster and falling towards it.

The emission from the radio galaxy B1610-60.5 (Jones \& Mc Adam 1992) has
been drawn into the right panel of Fig.~3 as well. It is one of the longest 
known head-tail galaxies (26 armin, $\sim 710$ kpc). The radio emission aligns
over nearly its full extent with the 3rd contour of the main X-ray 
component. The lack of distortion of the radiolobes of both radio
galaxies and the compactness of the subclump imply that the suspected
merger must be in an early stage. Forthcoming ATCA HI-synthesis observations 
will allow deeper insight into this cluster.
 
The cluster A3627 most likely is the previously unidentified core of
the Great Attractor overdensity. The mass excess of the GA is presumed
to arise within an area of radius of about 20-30 Mpc (Lynden-Bell
\etal 1988).
This actually matches the emerging picture from our observations 
quite well. A3627 seems the center of an apparent ``great wall''-like
structure, similar to Coma in the (northern) Great Wall:
a broad filament reaches from
$(\ell,b,v)=(335\deg,-25\deg,\sim 4500$ \kms) 
over the GA region towards ($295\deg,+5\deg,\sim 5500$ \kms). Whether 
it merges into the Vela-supercluster at
($280\deg,+6\deg,\sim 6000$ \kms) is not yet certain.

\section{Prospects for the Multi Beam ZOA-Survey}
The optical deep galaxy survey has reduced the gap of the ZOA and 
revealed many interesting extragalactic large-scale
structures. However, the innermost part of the southern Milky Way 
remains fully obscured. Here, the MB ZOA-survey will at last allow 
a view of that area of the extragalactic sky.

The effectiveness of \hi-observations of partially 
obscured galaxies at low latitudes substantiate the high expectations for
the MB ZOA-survey. This survey will trace the above discovered
structures in full detail - and also delineate the nearby, dynamically
important voids. The Hydra-Antlia extension can be followed even across
the inner part of the ZOA. An obscured dominant component might still 
reveal this structure to be a supercluster. 
The little studied Vela supercluster can be 
traced. The mapping of the central part of the GA can be completed,
revealing whether it merges with the Vela supercluster or bends back 
towards the Centaurus clusters. The extent and mass of the nearby
Puppis cluster ($245\deg,0\deg,\sim 1500$ \kms, Lahav \etal 1993) 
can be assessed and the extent of the Ophiuchus cluster close to the 
Galactic bulge at $\sim 8000$\kms\ (Wakamatsu \etal 1996) and its 
connections to other supercluster traced. Other important structures, 
not yet suggested from optical surveys can be mapped in the deepest layers
of obscuration.

These revelations all will lead to a better understanding of the 
galaxy distribution, the underlying mass distribution and the
dynamics of the local Universe.

\section*{Acknowledgments}
The research by RCKK is being supported with an EC-grant.
Financial support was provided by CNRS through the Cosmology GDR program. 
PAW is supported by the South African FRD. We particularly would like
to express our thanks to the staff of the Parkes Telescope for
their efficient support and their hospitality and 
look forward to working with the staff in the future. 
  
\medskip
\reference 
Abell, G.O., Corwin, H.G. \& Olowin, R.P. 1989, ApJSS 70, 1
\reference 
B\"ohringer, H., Neumann, D.M., Schindler, S. \& Kraan-Korteweg, R.C.
1996, ApJ, in press
\reference 
Burstein, D. \& Heiles, C. 1982, AJ 87, 1165
\reference 
Hoffman, Y. 1994,
in {\it Cosmic Velocity Fields}, 9$^{th}$ IAP Astrophysics Meeting, 
eds. F. Bouchet \& M. Lachi\`eze-Rey, Editions Frontieres, 357  
\reference
Jahoda, K. \& Mushotzky, R.F. 1989, ApJ 346, 638
\reference
Jones, P.A. \& McAdam, W.B. 1992, ApJSS 80, 137
\reference 
Kerr, F.J., Bowers, P.F., Jackson, P.D. \& Kerr, M. 1986, AASS 66, 373
\reference 
Kolatt, T., Dekel, A., \& Lahav, O. 1995, MNRAS 275, 797
\reference
Kraan-Korteweg, R.C. 1989, in {\it Rev. in Modern Astron.} 2, ed. G. Klare,
Springer: Berlin, 119
\reference
Kraan-Korteweg, R.C. \& Woudt, P.A. 1994,
in {\it Unveiling Large-Scale Structures behind the Milky Way},
4$^{th}$ DAEC Meeting, eds.
C. Balkowski \& R.C. Kraan-Korteweg, A.S.P. 67, p89
\reference
Kraan-Korteweg, R.C., Cayatte, V., Fairall, A.P., Balkowski, C. \&
Henning. P.A. 1994,
in {\it Unveiling Large-Scale Structures behind the Milky Way},
4$^{th}$ DAEC Meeting, eds.
C. Balkowski \& R.C. Kraan-Korteweg, A.S.P. 67, p99
\reference
Kraan-Korteweg, R.C., Woudt, P.A., Cayatte, V., Fairall, A.P.,
Balkowksi, C. \& Henning, P.A. 1996, Nature, 379, 519
\reference
Lahav, O., Edge, A.C., Fabian, A.C. \& Putney, A. 1989,
MRNRAS 238, 881
\reference
Lahav, O., Yamada, T., Scharf, C. \& Kraan-Korteweg, R.C. 1993,
MNRAS 262, 711
\reference
Lauberts, A. 1982 {\it The ESO/Uppsala Survey of the ESO (B) Atlas}, 
ESO: Garching
\reference
Lynden-Bell, D., Faber, S.M., Burstein, D., Davies, R.L., Dressler, A.,
Terlevich, R.J. \& Wegner, G. 1988, ApJ 326, 19
\reference 
Peebles, P.J.E. 1994, ApJ 429, 43
\reference 
Saunders, W., Frenk, C., Rowan-Robinson, M., \etal,
1991, Nature, 349, 32
\reference
Wakamatsu, K., Malkan, M., Parkes, Q.A. \& Karoji,H. 1996, these proceedings
\reference
Woudt, P.A., Fairall, A.P., Kraan-Korteweg, R.C. \& Cayatte, V.,
1996, in prep.
\end{document}